\documentclass[twocolumn,aps,pra,showpacs,amsmath,amssymb,nobibnotes,superscriptaddress]{revtex4-1}

\usepackage{graphicx}
\usepackage{amsmath,amssymb}
\usepackage{hyperref}

\begin{document}

\title{Heisenberg scaling of imaging resolution by coherent enhancement}

\author{Robert McConnell}
\email{robert.mcconnell@ll.mit.edu}
\affiliation{Lincoln Laboratory, Massachusetts Institute of Technology, Lexington, Massachusetts 02420, USA}

\author{Guang Hao Low}

\author{Theodore J. Yoder}
\affiliation{Massachusetts Institute of Technology, Cambridge, Massachusetts 02139, USA}

\author{Colin D. Bruzewicz}
\affiliation{Lincoln Laboratory, Massachusetts Institute of Technology, Lexington, Massachusetts 02420, USA}

\author{Isaac L. Chuang}
\affiliation{Massachusetts Institute of Technology, Cambridge, Massachusetts 02139, USA}

\author{John Chiaverini}

\author{Jeremy M. Sage}
\affiliation{Lincoln Laboratory, Massachusetts Institute of Technology, Lexington, Massachusetts 02420, USA}

\date{\today}

\begin{abstract}
Classical imaging works by scattering photons from an object to be imaged, and achieves resolution scaling as $1/\sqrt{t}$, with $t$ the imaging time. By contrast, the laws of quantum mechanics allow one to utilize quantum coherence to obtain imaging resolution that can scale as quickly as $1/t$ -- the so-called ``Heisenberg limit.'' However, ambiguities in the obtained signal often preclude taking full advantage of this quantum enhancement, while imaging techniques designed to be unambiguous often lose this optimal Heisenberg scaling. Here, we demonstrate an imaging technique which combines unambiguous detection of the target with Heisenberg scaling of the resolution. We also demonstrate a binary search algorithm which can efficiently locate a coherent target using the technique, resolving a target trapped ion to within 0.3\% of the $1/e^2$ diameter of the excitation beam.
\end{abstract}

\maketitle

\section{Introduction}

Imaging is an essential task in many areas of science, from biology to astronomy to condensed matter physics. Classically, imaging is performed by illuminating a target and collecting those photons which scatter from it. The scale of imaging resolution in this case is set by the wavelength $\lambda$ of the imaging light and the numerical aperture of the imaging system, but the resolution improves only as the square root of the number of scattered photons and hence as the square root of imaging time $t$. Especially in situations when the numerical aperture of the imaging system is limited, the practically-achievable resolution can be insufficient to resolve details of interest.

Numerous imaging techniques which outperform the traditional diffraction limit have been demonstrated \cite{HofmannRESOLFT2005,RustSTORM2006,BetzigPALM2006,TrifonovSTED2013}. These techniques typically work by ``excluding'' targets not within a sub-diffraction-limited area by storing such targets in a non-scattering, ``dark'' state $| d \rangle$, then scattering imaging photons off of remaining targets in a bright state $|g \rangle$ on a strong transition $|g \rangle \rightarrow |e \rangle$. While these techniques have realized resolution as low as $\lambda / 10$ in some cases \cite{BetzigSubDiffraction1992, HellNanoscopy2007}, they are still limited to a classical time scaling of $1/ \sqrt{t}$ because they do not utilize the full quantum mechanical coherence of their targets.

By contrast, the coherent properties of a quantum mechanical two-level system in principle allow resolution scaling at the so-called Heisenberg limit, as $1/t$ \cite{GiovannettiQuantumEnhancement2004,ElusiveHeisenberg2012}. A single spin precessing under a Hamiltonian $H$  accumulates phase $\phi = (H/\hbar) t$ and a single measurement of this phase achieves resolution $\Delta \phi = \pi / 2$ \cite{RamseySOF}. If the particle position $x$ can be linearly mapped to $H$, positional uncertainty $\Delta x \sim \Delta \phi /t$ scaling as $1/t$ can be achieved. 

One way to map particle position $x$ to $H$ is to utilize the spatially-varying intensity of a Gaussian beam coupling $|g \rangle$ to $|e \rangle$ and drive a single long pulse. In this case, it is total rotation angle on the Bloch sphere, $\theta$, which carries information about the particle position. However, when a single long pulse is used (Figure 1(a)), other rotation angles separated by a multiple of $2 \pi$ from the correct $\theta$ can lead to identical observables, which can render a precise estimation of the actual phase impossible. Techniques to produce unambiguous phase mappings \cite{VitanovNarrow2011, JonesNarrow2013,LowNarrow2014} often do so at the expense of Heisenberg scaling, returning to a classical scaling $\sim 1/ \sqrt{t}$. 

\begin{figure}
\label{QICompare}
\includegraphics[width=0.45 \textwidth]{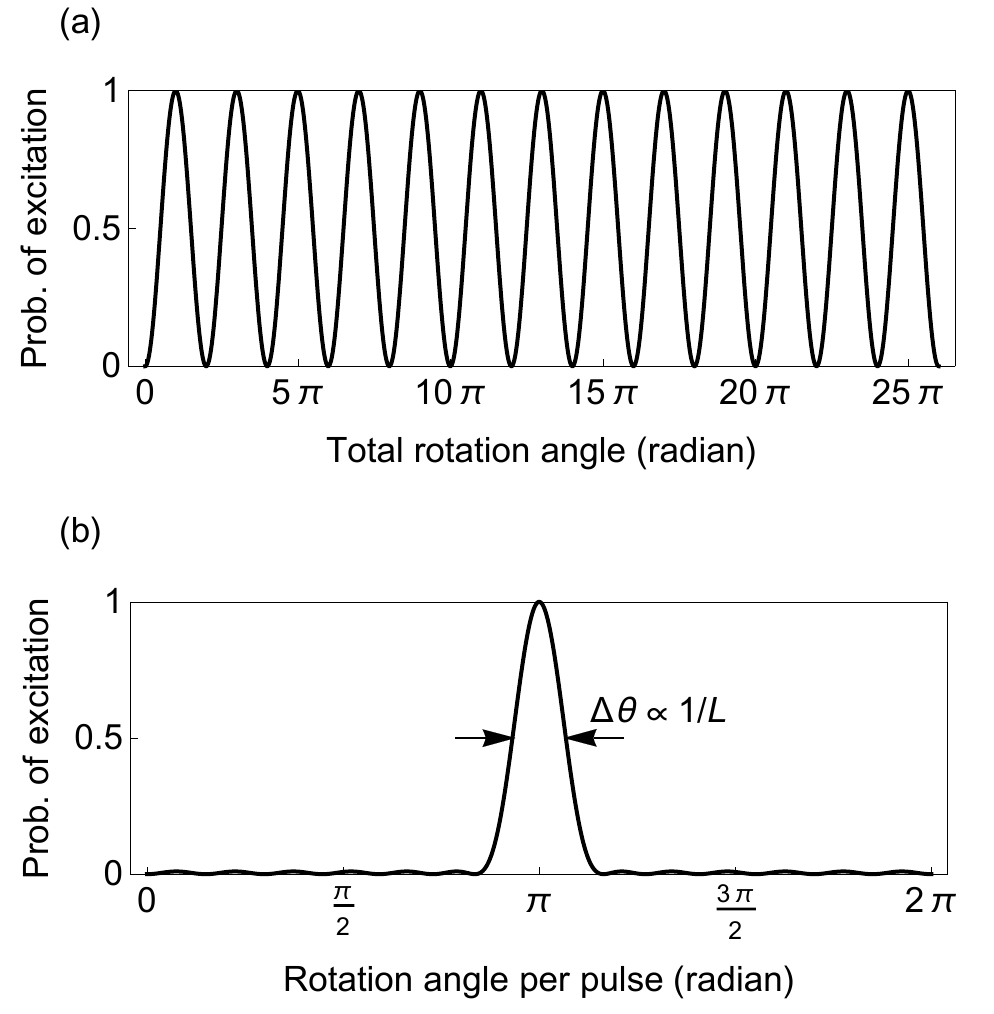}
\caption{(a) Coherent oscillations in a driven two level system produce very narrow features which in principle can be used to achieve Heisenberg-limited resolution, but signal ambiguities often preclude achieving such resolution in practice. (b) A sequence of $L$ pulses of the same overall duration as the drive in (a) can produce an unambiguous single excitation peak whose width nevertheless preserves optimal Heisenberg scaling as $1/L$, by varying the phases $\phi_i$ of each pulse. Here, $L = 13$.}
\end{figure}

In this proof-of-principle experiment, we demonstrate an imaging technique which exploits quantum mechanical coherence and optimal quantum control to unambiguously resolve a trapped atomic ion with Heisenberg scaling of the resolution. Optimally designed pulse sequences \cite{LowQuantumImage2015} transfer the target ion from its ground state $|g \rangle$ to excited state $|e \rangle$ with a position uncertainty scaling as $1/L$, with $L$ the length of the pulse sequence (Figure 1(b)). By using a binary search algorithm which starts with broad excitation and progressively narrows, we efficiently determine the ion position to within $0.3 \%$ of the control beam diameter, with search failure exponentially suppressed in the number of sequence repetitions. To our knowledge, this represents the first experiment to approach Heisenberg-limited scaling in an imaging task. The technique is related to methods that have achieved magnetic field resolution improving as $t^{-0.85}$ \cite{WaldherrHeisenbergMagnetometry2012,PuentesEfficientMagnetometry2014} and have improved the frequency stability of local oscillators \cite{SastrawanLOImprove2016}. High-resolution magnetic resonance imaging of a single diamond-NV center within an external magnetic gradient has been achieved via a Fourier-transform technique \cite{Arai15}. In contrast to those phase-estimation techniques, the method we employ here directly obtains positional information via the spatial gradient of the coherent drive's intensity. As a result, we do not require external fields for imaging (other than the coherent control drive) and we achieve rapid imaging of a coherent target in real time, with minimal post-processing of data. This allows achievement of a given resolution much faster than is possible classically, useful in any situation where the available time for imaging is limited. The narrow excitation window as a function of control drive intensity also allows this technique to be used for site-selective addressing of one ion or other coherent target within an array \cite{Wineland1998,ShenSiteAddress2013,MerrillPulseSequence2014}, and allows straightforward generalization to imaging of multiple targets.

Our quantum-enhanced imaging technique requires a coherent drive coupling the target states $|g\rangle, |e\rangle$ with a Rabi frequency $\Omega$ and able to implement arbitrary rotations on the Bloch sphere of an angle $\theta = \Omega t$. Quantum-enhanced imaging is implemented by a sequence of $L$ such rotations. Within such a sequence, each pulse is performed for the same time $t_0$ and with the same laser intensity such that the rotation angle $\theta$ per pulse is the same. However, the phase $\phi_i$ of each pulse $i$ in the sequence is optimized such that the ion is only transferred from $|g \rangle$ to $|e \rangle$ if the rotation per pulse satisfies $\theta = \pi$ to within an error $\Delta \theta \sim 1/L$ (Figure 1(b)). We here consider a 1-D case, but this technique can readily be generalized to higher dimensions. 

The Heisenberg-limited scaling of the error in rotation angle $\theta$ can be mapped to position resolution by using the spatially-varying intensity of a Gaussian beam. In particular, for a beam centered at the origin, the ion Rabi frequency as a function of position $x$  obeys
\begin{equation}
\Omega (x) = \Omega_0 e^{-x^2/w^2},
\end{equation}
where $w$ is the beam waist ($1/e^2$ intensity radius) and $\Omega_0$ the ion Rabi frequency at the center of the beam. The positional mapping is optimized if the excitation occurs at the location of maximum field slope of the beam, that is, if $\Omega(x) t_0= \pi$ for the point where $| d \Omega / d x|$ is maximized. This maximum occurs when the beam intensity and pulse length $t_0$ are chosen such that $\Omega_0 t_0 = \pi \sqrt{e}$. For this $\Omega_0$, and at the location of maximum slope $x = w/\sqrt{2}$, the positional error $\Delta x$ obeys
\begin{equation}
\Delta x = \frac{\Delta \theta \cdot w}{\pi \sqrt{2}}.
\end{equation}
By correctly choosing the pulse duration as a function of Rabi frequency, this optimal positional resolution can be retained even as the control beam is scanned to search for the ion location.

The phases that produce such a narrowband excitation are described in \cite{LowQuantumImage2015, LowOB12016} and are derived from Chebyshev polynomials. In essence, these pulse sequences trade small probabilities of excitation (``ripples'') in the stopband for an optimally narrow passband, in analogy to Chebyshev filters (c.f. Figure 1(b)).

\section{Apparatus}

\begin{figure}
\label{Apparatus}
\includegraphics[width=0.5 \textwidth]{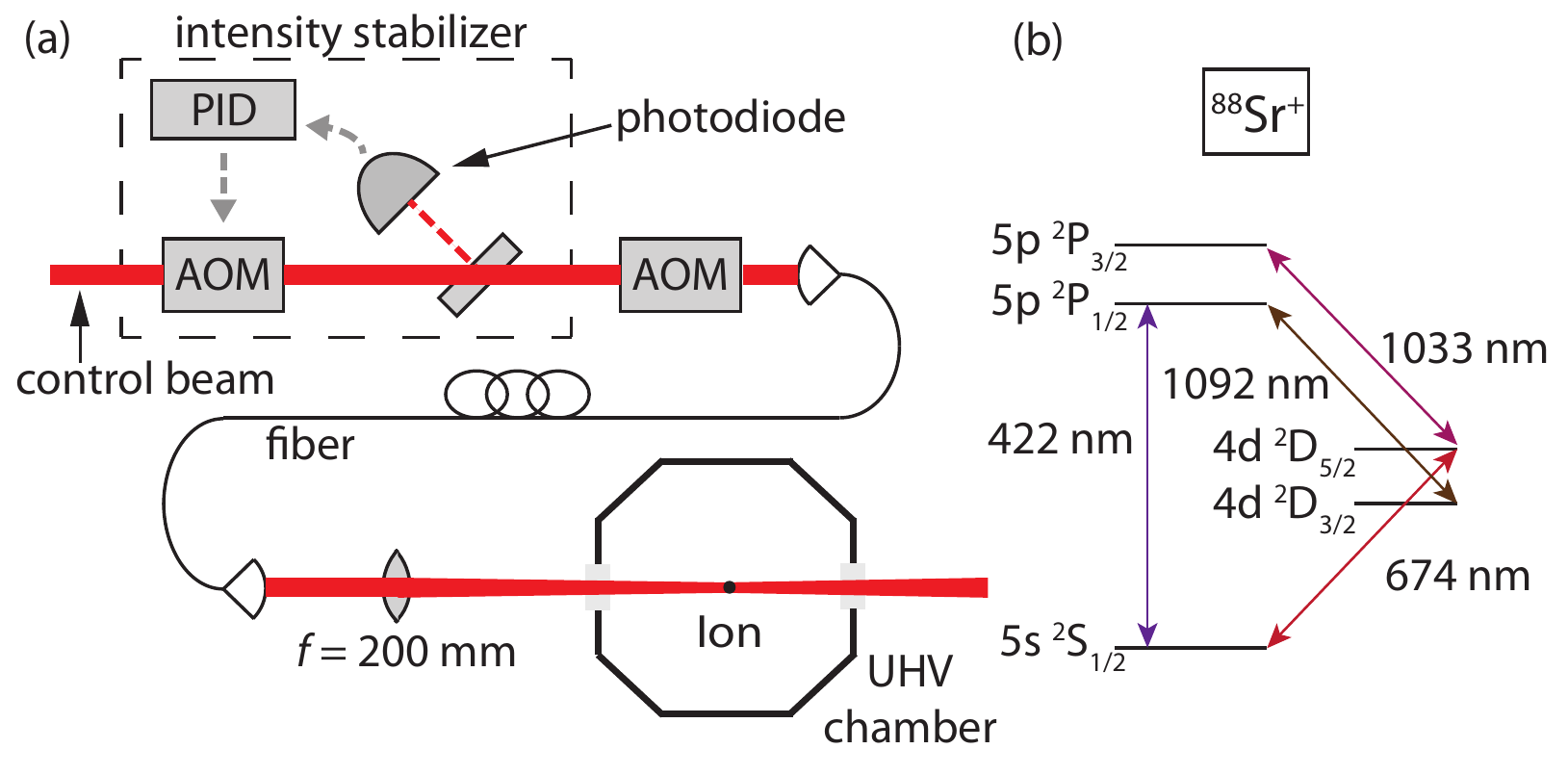}
\caption{(a) Schematic of the apparatus used to perform quantum-enhanced imaging. The 674 nm control beam passes through an intensity stabilization circuit (described in text) and then an AOM used to control its intensity. The beam passes through a fiber and then enters the apparatus, where it is focused onto the ion. PID = proportional-integral-differential feedback controller. (b) Relevant level structure in $^{88}$Sr$^{+}$.}
\end{figure}

Figure 2 shows the apparatus we use to perform quantum-enhanced imaging. A trapped $^{88}$Sr$^{+}$ ion is confined 50 $\mu $m above the surface of a niobium surface electrode trap and is laser-cooled to its motional ground state \cite{sage2012loading}. The ion is 
coherently driven on the 674 nm, $| g \rangle = |5 S_{1/2}, m= -1/2 \rangle \rightarrow | e \rangle = |4 D_{5/2}, m = -5/2 \rangle$ quadrupole 
transition with lifetime $\sim$ 0.5 s. Approximately 3 mW of power from a diode laser provide a Rabi frequency of typically $\Omega_0 = 2 \pi \times 100$ kHz. This beam is stabilized by transmission through a narrow-linewidth ultra-low-expansion (ULE) glass cavity \cite{AkermanGates2015}, with the transmitted beam seeding an injection-locked laser which is then amplified by a tapered amplifier. This method filters out spectral noise (``servo bumps'') in the laser which would otherwise cause degradation of the technique. An intensity stabilization circuit is used to limit intensity fluctuations of the control laser at the ion location. A photodiode samples a portion of the beam power; this photodiode signal is sent to a proportional-integral-differential feedback controller which adjusts the modulation input of an acousto-optical modulator (AOM) to maintain constant power. After the intensity stabilizer, we use a second AOM to control the power and phase of the control beam at the ion location. Finally, this output passes through a single-mode fiber and emerges from a fiber launch near the experiment which minimizes angular jitter of the beam. This output is focused on the ion by a lens of focal length $200$ mm. By adjusting the lens position with a manual micrometer and measuring the change in ion Rabi frequency with position, we determine the control laser beam waist to be $140 \pm 10 \, \mu$m at the ion location. These measurements also confirm the Gaussian shape of the control beam.

High-fidelity readout of the ion's internal state is accomplished by scattering light from a 422 nm laser which couples $|g \rangle$ but not $|e \rangle$ to the short-lived excited state $5 P_{1/2}$; a high-NA lens and external PMT collect scattered photons and allow readout fidelity of 99.99$\%$ in 1 ms. Additional repumping lasers at 1092 nm and 1033 nm used at various times during the control and readout sequence prevent population trapping in undesired internal states of the ion. 

\section{Results}

\begin{figure}
\label{ExampleFig}
\includegraphics[width=0.5 \textwidth]{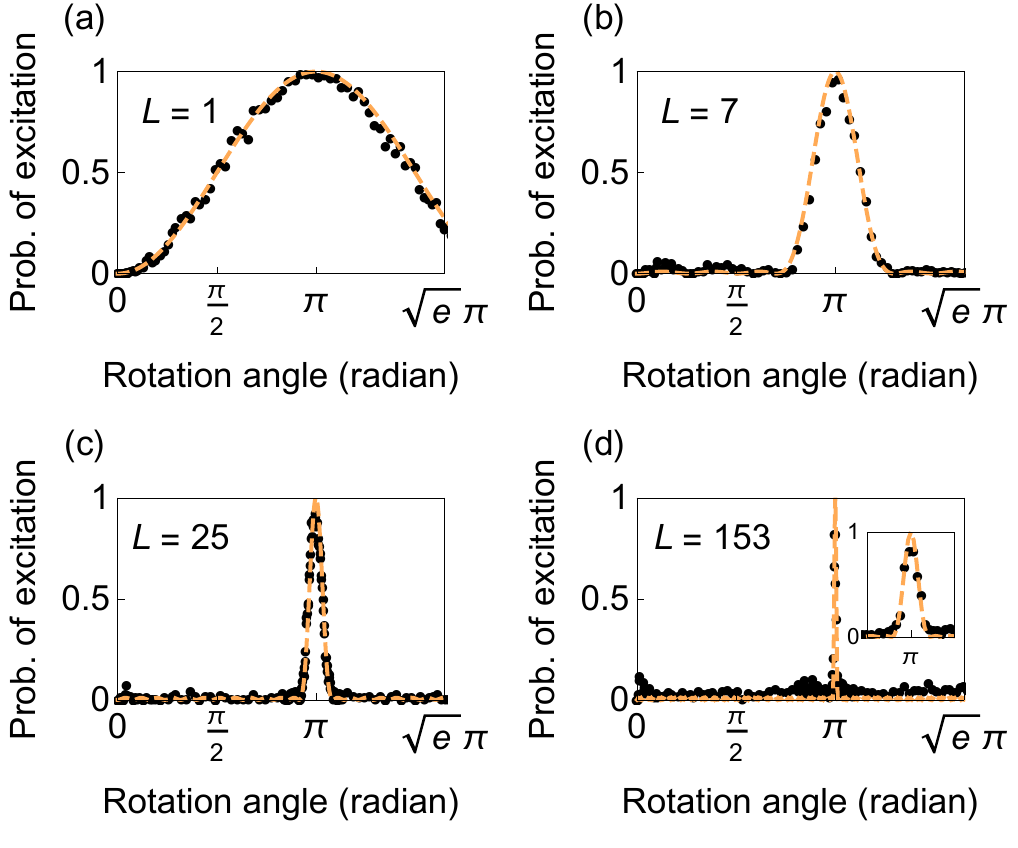}
\caption{Examples of quantum-enhanced imaging for (a) a single pulse, (b) a 7-pulse sequence, (c) a 25-pulse sequence, (d) a 153-pulse sequence. The orange dashed curves show model-free theory predictions for the excitation as a function of drive strength. The inset in subplot (d) shows the excitation width and agreement with the model near $\theta = \pi$ for the narrow $L = 153$ sequence; the inset x-axis spans 0.2 radian.}
\end{figure}

Figure 3 shows the results of applying excitation pulse sequences of varying length to the trapped ion. With the ion initially in $|g \rangle$, a pulse sequence of length $L$ and per-pulse rotation angle $\theta$ is applied; afterwards, we measure the final state of the ion. We perform 200 repetitions per point in order to estimate the probability of the transition $|g \rangle \rightarrow |e \rangle$ for each $L$ and $\theta$. For the experiments shown in Figure 3, the ion is located at the center of the beam and the beam intensity adjusted via the control AOM. Figure 3(a) shows a single pulse applied to the ion, in which case a broad excitation $|g \rangle \rightarrow |e \rangle$ occurs which does not well localize the ion. As the number of pulses is increased (Figure 3(b)-(d)), a narrow and unambiguous excitation is achieved. The dashed curves in the figure show the theoretical transfer probability; we demonstrate excitation widths that are in very good agreement with the theoretical predictions.

\begin{figure}
\label{ScalingFig}
\includegraphics[width=0.5 \textwidth]{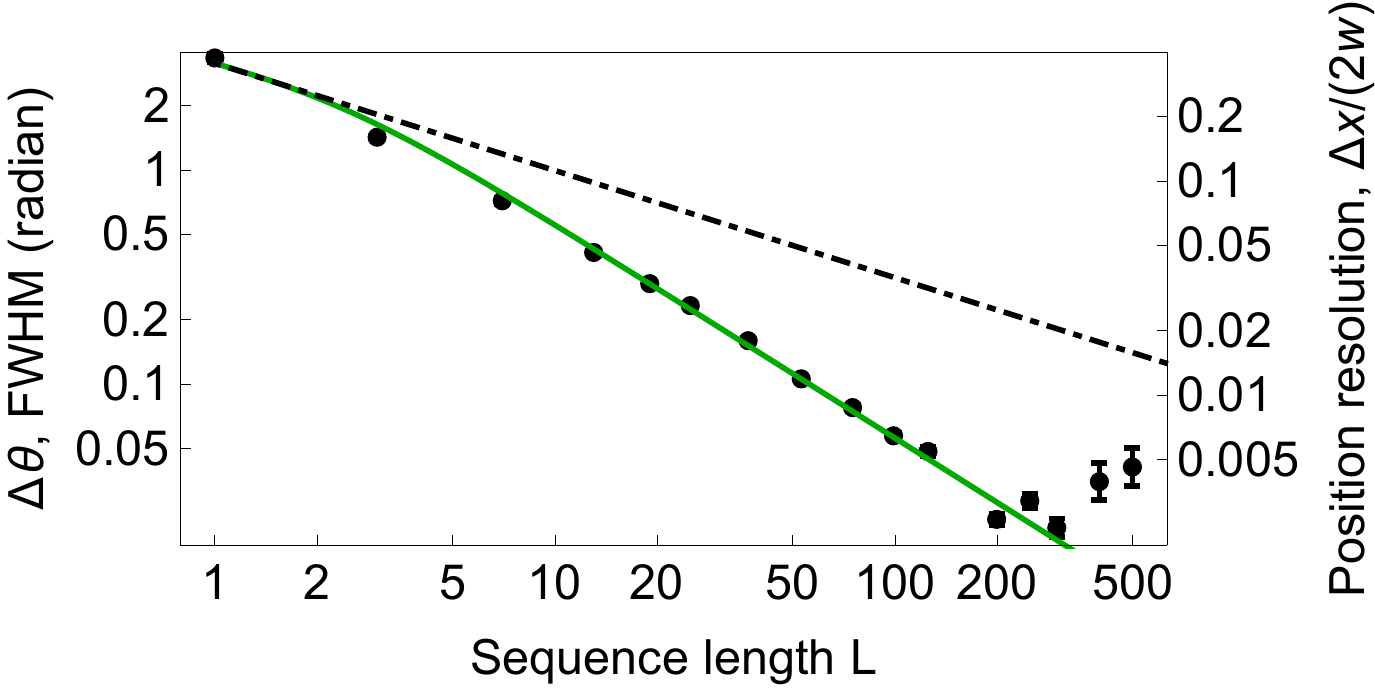}
\caption{Scaling of the FWHM of the excitation region as a function of pulse sequence length $L$. The solid green curve shows the theoretical width that should be achieved (asymptotically scaling as $1/L$),while the black dot-dashed curve shows classical scaling as $L^{-0.5}$.}
\end{figure}

Figure 4 shows the scaling of the fitted full width half maximum (FWHM) of the peaks in the experimental data as a function of sequence length $L$. The green curve shows the theoretical prediction with asymptotic $L^{-1}$ scaling, while the classical $L^{-0.5}$ scaling is shown by the black dash-dotted curve. Up to $L = 199$ we observe widths in very good agreement with the theory curve, while for $L > 199$ the width becomes limited by the finite system coherence time. We note that with per-pulse times of $10 \, \mu$s, we are able to perform pulse sequences somewhat longer than the nominal system coherence time of $\sim \! \! 0.5$ ms given by Ramsey measurements, due to partial spin-echo effects during the pulse sequences.

\begin{figure}
\label{LogSearchFig}
\includegraphics[width=0.5 \textwidth]{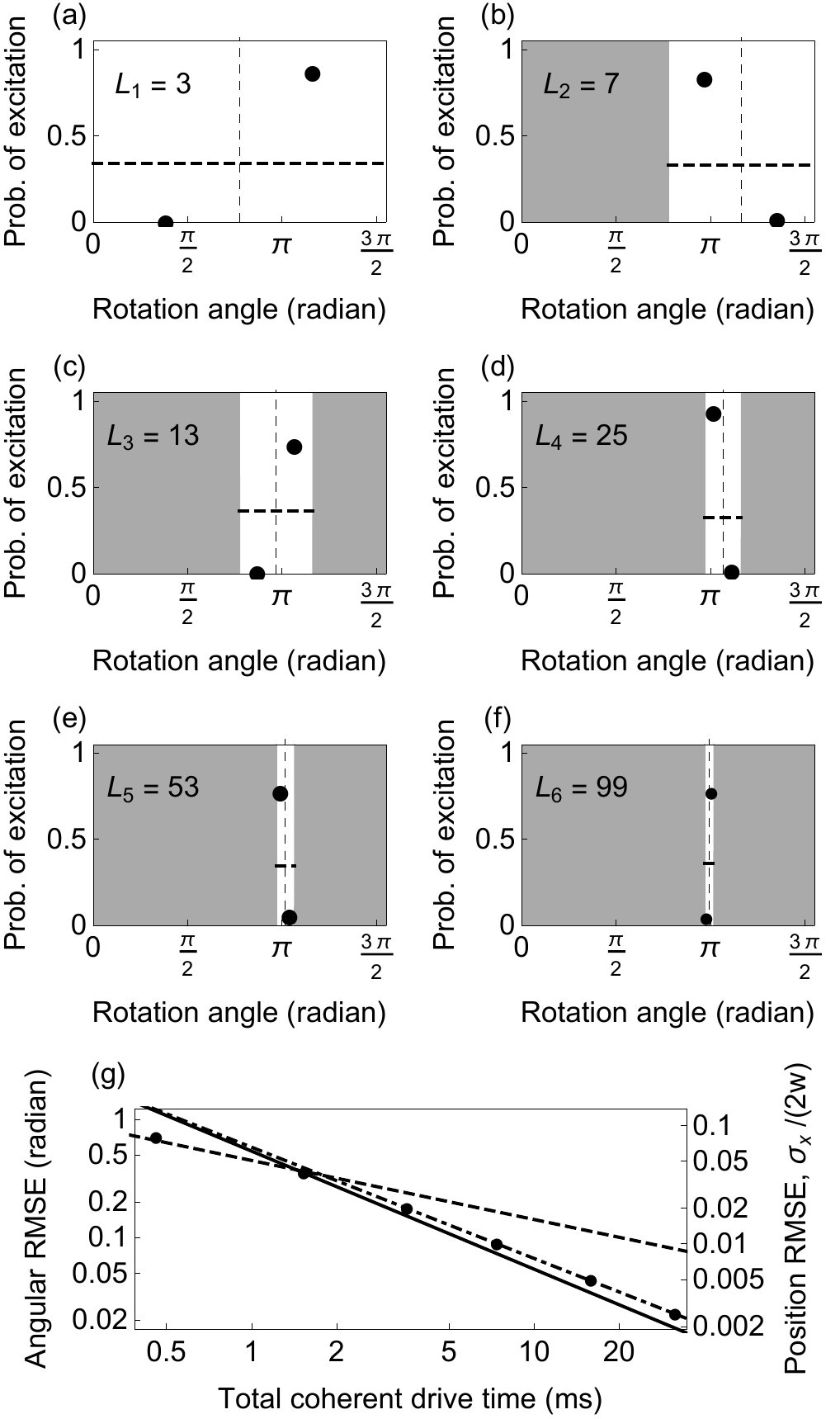}
\caption{Demonstration of the logarithmic search technique. The beam is positioned so that its maximum intensity slope is located at the ion position. (a) - (f) Successive pulse sequences of length $L_1 = 3$, $L_2 = 7$, $L_3 = 13$, $L_4= 25$, $L_5 = 53$, $L_6 = 99$ pulses are used to excite the ion. The amplitude is varied to determine the regions where the ion is excited. For iterations $i = 1...6$, thresholds $T_i $ ranging from 0.32 to 0.36 (dashed horizontal lines) are used to maximize the probability of success (Supplemental Information). Black dots indicate the average excitation probability observed when searching each of the two amplitude subspaces. Only $n = 5$ repetitions need be used for each scan due to the unambiguous, narrow feature inherent to the excitation sequences. With each iteration, searched locations where the ion was not found can be excluded (gray regions), the ion position is further localized, and subsequent pulse sequences search only in the remaining target area. (g) RMS error of the quantum-enhanced technique (filled circles) as a function of total coherent drive time clearly demonstrates the asymptotic $1/\tau$ scaling with total coherent drive time $\tau$ (solid curve). The dot-dashed curve is a fit showing scaling as $\tau^{-0.94 \pm 0.01}$, improving much more quickly than the classical $t^{-0.5}$ scaling (dashed curve).}
\end{figure}

Finally, we demonstrate a binary search technique to locate an ion of an initially unknown location (Figure 5). The essence of the binary search is as follows. An ion of initially unknown location is addressed by the coherent control beam. For convenience we assume the ion is in the half-space $x>0$ for which the ion Rabi frequency $\Omega(x) \propto e^{-x^2/w^2}$ is monotonically decreasing; thus, an unambiguous mapping between position $x$ and local Rabi frequency $\Omega(x)$ exists. (Relaxing this requirement would necessitate only one additional, short measurement to determine whether the ion was in the positive or negative half-space of the beam and would thus have negligible impact on the total imaging time.) A short sequence of length $L_1$ is initially applied to the ion. The initial search space is divided into a small number $M$ of different locations identified by position $x_k, k=1...M$; at each location $x_k$ we perform $n$ repetitions of the pulse sequence to attempt to drive the ion to $| e \rangle$. If the fraction of successful excitations of the ion out of $n$ repetitions exceeds a user-specified threshold $T$ at one of the $M$ locations and at no others, the ion is considered to be found at that location. This localizes the ion to the position $x_k$ to within an error $\Delta x_1  \propto 1/L_1$. The length of the sequence is then increased to $L_2$, and the resulting subspace of size $\Delta x_1$ is then itself divided into $M$ search locations. This process is repeated, each iteration $i$ localizing the ion to a corresponding $\Delta x_i \propto 1/L_i$ until the final $\Delta x_f$ (achieved with sequence length $L_f$) is less than some specified uncertainty goal. We note that the total time for binary search scales as $M n L_f$ and achieves positional error $\propto 1/L_f$, thus still retaining Heisenberg scaling.

Rather than physically moving our control beam, we instead vary its intensity and search for the amplitude $\Omega_{target}$ such that $\Omega_{target} e^{-x^2/w^2} t_0 = \pi$. We choose the ion position $x = w/\sqrt{2}$ to maximize the intensity gradient, and  hence positional resolution, of our beam. The results are shown in Figure 5, plotting the achieved resolution versus total coherent illumination time. We use $M = 2$ searches per iteration of the sequence, $n = 5$ repetitions per search location, and proceed from the $L_1 = 3$ pulse sequence to the $L_6 = 99$ pulse sequence to localize the ion. The threshold $T_i$ is optimized for each sequence (Supplemental Information).  The entire binary search algorithm was repeated 100 times to accumulate statistics. A failure of the binary search algorithm, in which case the algorithm fails to identify the correct subspace in which the particle is located (see Supplemental Information for more details), occurred only once during the 100 trials, and only in the final iteration $i = 6$. After six iterations of this procedure, and taking into account the finite chance of failure, the ion is localized to within a rotation angle root-mean-square error (RMSE) $(\sigma_{\theta})/(2 \pi) = 0.004$. Given our beam waist of $140 \pm 10 \, \mu$m, the spatial resolution we achieve via binary search corresponds to positional RMS uncertainty $\sigma_{x} = 0.7 \pm 0.1 \, \mu$m, or an uncertainty of $(0.3 \pm 0.03) \%$ of the $1/e^2$ beam diameter of $280 \, \mu$m. Figure $5$(g) shows the RMS error for the entire search taking into account the possibility of failure in a particular iteration, demonstrating resolution scaling with total coherent drive time $\tau$ as $\tau^{-0.94 \pm 0.01}$. 

Comparison of quantum-enhanced imaging to classical imaging is not straightforward as the two techniques do not use the same laser beam (the coherent 674 nm beam is used for quantum-enhanced imaging, while the 422 nm beam is used for classical imaging by scattering) and any comparison depends upon detailed experimental parameters. For our system, the quantum-enhanced technique exceeds the resolution achievable by classical scattering measurements, given the same total illumination time, for five or more iterations of the binary search (Supplemental Information). Residual slow beam positional fluctuations prevented further iterations of the binary search in this experiment. We note that, for these experiments, total experimental time is dominated by our relatively slow readout of $\sim 1$ ms per iteration, but much faster ion readouts (less than 150 $\mu$s) have been demonstrated \cite{MyersonHighFReadout2008}. This could possibly be implemented in our setup via adaptive measurement techniques or integrated detectors; we are currently pursuing efforts to demonstrate on-chip integrated detectors which could potentially enable much faster readout.

\section{Discussion}

In conclusion, we have demonstrated a new method of imaging which takes advantage of quantum coherence to achieve Heisenberg-limited scaling of the resolution. In our proof-of-principle experiment, we have achieved positional resolution to within $0.3 \%$ of the beam diameter and achieve spatial resolution of 0.7 $\mu$m, comparable to the probe laser wavelength of 674 nm. This is, to our knowledge, the first experiment to approach Heisenberg scaling of the resolution in an imaging task. 

The technique has many possible uses in traditional imaging applications, especially for longer wavelengths (e.g., microwaves) where the size of the imaging beam may be large compared to details of interest. Furthermore, the selective excitation provided by these pulse sequences may be useful for site-selective control of trapped ions or other quantum systems, such as quantum dots or nitrogen vacancy centers. As one example, a long pulse sequence---with correspondingly high spatial selectivity---may be used to drive a target rotation in a particular qubit in an array while minimizing crosstalk effects on other nearby qubits. Our finite coherence time, limited by non-Markovian magnetic field drifts and laser phase noise, as well as residual beam jitter prevent us from achieving sub-wavelength resolution in this proof-of-principle experiment; however, modest improvements to the system would enable us to retain Heisenberg scaling out to many hundreds of pulses and achieve such resolution. Though Heisenberg scaling holds only within the system coherence time, after which classical scaling again takes over, the use of this technique can achieve a given resolution much faster than classical imaging methods. 

Finally, while our proof-of-principle demonstration here images only one ion, the technique can straighforwardly be generalized to multiple targets within the search region, as long as the state detection allows determination of the total number of targets that have been excited---true for our trapped-ion detection here and for most other detection schemes based on fluorescence collection. In this case, each iteration of the search investigates any regions of the search space known to contain at least one target, with sub-regions not containing a target excluded from the later iterations (Supplemental Information).

\section{Acknowledgements}

We thank Peter Murphy and Chris Thoummaraj for assistance with ion trap chip packaging. This work was sponsored by the National Reconnaissance Office (NRO) and performed under Air Force Contract \#FA8721-05-C-0002. Opinions, interpretations, conclusions, and recommendations are those of the authors and are not necessarily endorsed by the United States Government.

\section{Supplemental Information}

\subsection{Binary search algorithm}

The binary search algorithm is used to locate a coherent target of initially unknown location. Here, we analyze the probability of success of this algorithm for realistic conditions.

The binary search algorithm uses a number $N$ of iterations, each of which divides the available search space into $M$ subspaces. In this analysis we focus on $M = 2$, although the results generalize to higher $M$. Here we work in the space of rotation-angle per pulse, which maps to positional space via the intensity gradient of the coherent beam. For a gven iteration $i$, the ion is located in one of the $M$ subspaces. $n_i$ repetitions of the pulse sequence are used to attempt to excite the ion in each of the subspaces. A threshold $T_i$ is specified for each iteration: if the fraction of successful excitations exceeds $T_i$, the ion is considered to be found within that subspace, which is then divided in two and searched again. For a given iteration, the search within the subspace where the ion is located can yield either a true positive (ion correctly found) or false negative (ion incorrectly not found). For the other subspace, the possibilities are false positive (ion incorrectly ``found'' when not present) and true negative (ion correctly not found). There are thus in total $2 \times 2$ possibilities. In the case of a true positive/true negative, the ion is correctly identified in one of the two subspaces and the search proceeds to the next iteration. In the case of false negative/true negative or true positive/false positive, the current iteration is inconclusive and must be repeated, but the failure of the current iteration is known and can be corrected. Only in the case of a false negative/false positive does the search fail in a way that leads to an incorrect final result. However, this probability is exponentially suppressed in the number of repetitions $n_i$.

For a given iteration, two subspaces are searched. For each of these subspaces, the control beam position is chosen so that the excitation region is in the middle of the subspace. Given a uniform prior distribution, and if the probability of excitation at each rotation angle $\theta$ is given by $f(\theta)$, the overall probability to excite if the ion is in subspace $S$ of width $\Delta \theta$ is 
\begin{equation}
p_S = \frac{1}{\Delta \theta} \int_{\theta \in S} f(\theta) d \theta.
\end{equation}
Similarly, the probability to excite if the ion is not in $S$ is 
\begin{equation}
p_{\sim S} = \frac{1}{\Delta \theta} \int_{\theta \notin S} f(\theta) d \theta.
\end{equation}
These probabilities can be used to find the overall chance that a given iteration of the binary search ultimately succeeds and how many repetitions of this iteration are needed. For a search in subspace $S$ with number of pulse sequence repetitions $n_i$ and threshold $T_i$, the chance that greater than $n_i T_i$ successes occur is given by the binomial distribution,
\begin{equation}
p_{S,n_i, T_i} = \sum_{k = \lceil n_i T_i \rceil}^{n_i} p_S^k (1-p_{\sim S})^{n_i-k} \frac{n_i!}{k! (n_i-k)!}.
\end{equation}
A similar expression describes the chance of $n_i T_i$ or more successes in the other search space $\sim$$S$. We denote the probability for a true positive/true negative as $p_{t,t}^{(i)}(n_i, T_i)$, the probability for a true positive/false positive as $p_{t,f}^{(i)}(n_i, T_i)$, and so on. These probabilities may be written as
\begin{widetext}
\begin{align}
p_{t,t}^{(i)}(n_i, T_i)&= p_{S,n,T} (1-p_{\sim S, n, T}) =  \sum_{k = \lceil nT \rceil}^{n}\frac{p_S^k (1-p_{\sim S})^{n-k} n!}{k! (n-k)!} \, \left(1-  \sum_{k = \lceil nT \rceil}^{n} \frac{p_{\sim S}^k (1-p_S)^{n-k} n!}{k! (n-k)!} \right)  \\
p_{t,f}^{(i)}(n_i, T_i)&= p_{S,n,T} p_{\sim S, n, T} =  \sum_{k = \lceil nT \rceil}^{n} \frac{ p_S^k (1-p_{\sim S})^{n-k} n!}{k! (n-k)!} \, \sum_{k = \lceil nT \rceil}^{n}  \frac{p_{\sim S}^k (1-p_S)^{n-k} n!}{k! (n-k)!}  \\
p_{f,t}^{(i)}(n_i, T_i)&= (1-p_{S,n,T}) (1-p_{\sim S, n, T})  =  \left( 1- \sum_{k = \lceil nT \rceil}^{n}  \frac{p_S^k (1-p_{\sim S})^{n-k} n!}{k! (n-k)!} \right) \left(1-  \sum_{k = \lceil nT \rceil}^{n} \frac{ p_{\sim S}^k (1-p_S)^{n-k} n!}{k! (n-k)!} \right)   \\
p_{f,f}^{(i)}(n_i, T_i)&= ( 1- p_{S,n,T}) p_{\sim S, n, T} =  \left( 1- \sum_{k = \lceil nT \rceil}^{n}  \frac{p_S^k (1-p_{\sim S})^{n-k} n!}{k! (n-k)!} \right)  \, \sum_{k = \lceil nT \rceil}^{n}  \frac{p_{\sim S}^k (1-p_S)^{n-k} n!}{k! (n-k)!}  
\end{align}
\end{widetext}
For $N$ iterations, the overall probability that the search succeeds is 
\begin{equation}
\prod_{i=1}^{N} (1-p_{f,f}^{(i)}(n_i, T_i)).
\end{equation}
The number of repetitions $n_i$ within iteration $i$ can be set so that the overall probability of failure is below a specified value. Note that $p_{f,f}^{(i)}(n_i,T_i)$ decreases exponentially with $n_i$, allowing the binary search to be efficient even in the case of imperfect pulses. For example, for $p_S = 0.7$, $p_{\sim S} = 0.1$, $N = 6$ iterations, and an overall failure rate of less than $1 \%$, the required number of repetitions $n = 5$.

The total coherent drive time, $\tau$, needed to complete a binary search is then
\begin{equation}
\tau = \frac{\pi \sqrt{e}}{\Omega_0} \sum_{i=1}^{N}2  L_i n_i \frac{(1- p_{f,f}^{(i)})}{p_{t,t}^{(i)}}.
\end{equation}
Since the probability $p_{t,t}^{(i)}$ is typically $0.7 - 0.8$, the number of times an iteration is repeated is relatively small and these repetitions have little effect on the overall time needed to complete the binary search.

The optimal threshold fraction $T_i$ must maximize the probability $p_{t,t}^{(i)}(n_i, T_i)$ that both a true-positive and true-negative result is obtained. This threshold can be found via maximum-likelihood analysis over the two binomial distributions that contribute to $p_{t,t}^{(i)}$, but is well approximated by $T \approx (p_{S} + p_{\sim S})/2$. For the actual implementation of the binary search shown in the text, the thresholds are found to be $T_1 = 0.34, T_2 = 0.33, T_3 = 0.36, T_4 = 0.32, T_5 = 0.35, T_6 = 0.36$. With these thresholds, 99 of 100 binary searches were completed successfully and the single failure occurred only at the final step $i = 6$.

To calculate the RMS error resulting from the binary search, we begin with the width $d \theta$ of a given iteration of the binary search $i$, expressed as $d \theta^{(i)} = 2^{-i} \sqrt{e} \pi$. Successful completion of this iteration of the binary search localizes the ion with RMS error $\sigma_{\theta}^{(i)} = d \theta^{(i)} / (2 \sqrt{3})$, given by integration of a uniform distribution over a segment of length $d \theta^{(i)}$. Eqn. (2) can then be used to map angular RMS error $\sigma_{\theta}$ to positional RMS error $\sigma_{x}$. To take into account the chance of failure at a given iteration, if the search fails at iteration $i$, then the particle is erroneously believed to be an angular distance $d \theta^{(i)}$ away, leading to RMS error $\sigma_{\theta}^{(i)} \approx d \theta^{(i)}$, and no further iterations will succeed. Taking into account the measured probabilities for failure at each step, the overall RMS error $\sigma_x$ can be accurately calculated.

\subsection{Detection of multiple targets}

While the imaging algorithm we have demonstrated in this paper is for a single target, the algorithm can also be generalized to multiple targets, as long as the readout allows discrimination between different numbers of targets that have been excited by the coherent drive. For fluorescence-based detection, this is generally possible as the total received fluorescence is proportional to the number of targets in $| e \rangle.$ Reliable discrimination of the number of targets in $|e \rangle$ requires that the signal-per-target exceeds the system noise. Assume that $C$ photons per target are collected in a system of $N$ targets with background (dark) counts $B$ within the experiment measurement time. If all counts follow a Poisson distribution, then the difference in signal between $N_e$ and $(N_e - 1)$ excited targets is $C$, while the noise will be $\approx \sqrt{N_e C + B}$. Since the observed counts must be within $\pm C/2$ of the expected value for reliable detection, and for $N_e C \gg B$, this requires that $\sqrt{C/2 N_e}$ be on the order of a few. For our trapped-ion system $ C \approx 100$ counts in 1 ms, which gives $N_e = 3$ or 4 ions in the search space (beam width) for signal-to-noise of a few.

If the total number of targets is known in advance, then the analysis from Section A applies. The only difference is that after a given iteration of the search, the algorithm proceeds if exactly the known number of targets have been found in the search space, while the iteration must be repeated if a different number is found (indicating that at least one false positive or false negative occurred). For the subsequent iteration, all subspaces of the search containing at least one target will be divided in two and re-searched; any subspaces not containing any targets can be ignored.

If the total number of targets to be located is not initially known, then the number of targets can be determined as a preliminary step in the algorithm. This is accomplished by use of a broadband pulse to excite all targets simultaneously with high probability, followed by fluorescence detection, with the number of targets within the beam determined by the total fluorescence collected. Simultaneous transfer of all targets to $| e \rangle$ can be accomplished with a pulse sequence optimized for broadband excitation \cite{LowOB12016} recently developed by some of us. Knowledge of the number of targets within the search space immediately allows quantum-enhanced imaging to proceed as per Section A. This preliminary step makes a negligible contribution to the total search time.

\subsection{Quantum speedup}

The net speedup provided by quantum-enhanced imaging, as compared to a classical imaging technique, depends upon detailed experimental parameters. The classical technique uses scattering from a strong transition, while the quantum-enhanced technique utilizes a coherent drive on a narrow-linewidth transition combined with occasional state readout on the strong transition. The two techniques are most easily compared by normalizing to the waist $w$ of the respective beams (classical imaging/readout beam and coherent beam). Classical imaging proceeds by collecting scattered photons to determine the ion position within the imaging beam, with error being given by the Poissonian noise on the number of collected photons. If the point of maximum slope of the classical imaging beam is used, classical imaging achieves RMS resolution 
\begin{equation}
\frac{\sigma_{x}^{(c)}}{w} = \frac{e^{1/4}}{2 \sqrt{\Gamma_{sc} \eta_c \tau}},
\end{equation}
with $\Gamma_{sc}$ the scattering rate of the readout transition, $\eta_c$ the overall collection efficiency of the detector, and $\tau$ the measurement time. By comparison, quantum-enhanced imaging for a maximum sequence length $L_{N}$ requires a total coherent drive time time given by
\begin{equation}
\tau =\sum_{i=1}^{N} \left(  \frac{\pi \sqrt{e}}{\Omega_0}  2 L_i n_i \frac{(1- p_{f,f}^{(i)})}{p_{t,t}^{(i)}} \right).
\end{equation}
Quantum-enhanced imaging for a sequence length $L_{N}$ yields RMS resolution equal to 
\begin{equation}
\frac{\sigma_x^{(q)} }{w} = \frac{0.36}{L_{N}},
\end{equation}
with 0.36 a numerical factor derived from Eqn. (2) and the RMS error analysis from Section A.

\begin{table*}
\begin{center}
\begin{tabular}{| c | c | c | c |}
\hline
Sequence length $L$ & Total illumination time (ms) & Resolution $\sigma_x^{(q)} /w$ & Achievable classical resolution $\sigma_x^{(c)} /w$ \\  \hline
3 & 0.5 & 0.16 & 0.071 \\ \hline
7 & 1.5 & 0.079 & 0.039 \\ \hline
13 & 3.6 & 0.040 & 0.026 \\ \hline
25 & 7.4 & 0.020 & 0.018 \\ \hline
53 & 15.9 & 0.0099 & 0.012 \\ \hline
99 & 31.6 & 0.0051 & 0.0086 \\ \hline
\hline
\end{tabular}
\caption{Comparison of classical and quantum-enhanced imaging resolution for the specific experimental parameters used in this demonstration.}
\end{center}
\end{table*}

A comparison of the achievable resolution via classical and quantum-enhanced imaging can therefore not be done fundamentally but requires knowledge of detailed experimental parameters. For our particular system parameters, we have included this analysis below in Table S. 1.

Table 1 demonstrates that, for our specific experimental parameters and errors, the final two iterations of the binary search with $L > 50$ exceed the resolution which can be achieved for the same illumination time via scattered light from a classical beam in our experiment.

\end{document}